\documentclass[12pt]{article}
\usepackage{amsmath}
\usepackage{graphicx,psfrag,epsf}
\usepackage{enumerate}
\usepackage{natbib}
\usepackage{makecell}
\usepackage[title]{appendix}
\setcitestyle{parentheses}
\usepackage{url} 
\usepackage[colorlinks,
            linkcolor=blue,
            anchorcolor=blue,
            citecolor=blue
            ]{hyperref}
\numberwithin{equation}{section}
\numberwithin{figure}{section}
\numberwithin{table}{section}

\usepackage{float}
\usepackage{amsthm,amssymb}
\usepackage{multirow,booktabs}
\usepackage{bbm}
\allowdisplaybreaks[4]

\DeclareMathOperator*{\argmin}{arg\,min}

\newcommand{\blind}{1}

\newtheorem{assumption}{Assumption}[section]

\newtheorem{definition}{Definition}[section]
\numberwithin{equation}{section}

\date{}

\begin{document}

\def\spacingset#1{\renewcommand{\baselinestretch}%
{#1}\small\normalsize} \spacingset{1}

\title{ \vspace{-1.5cm}  \bf The survival-incorporated median versus the median in the survivors or in the always-survivors: What are we measuring? And why?}

\author{Qingyan Xiang, \\ \vspace{0.cm}
    Department of Biostatistics, Boston University \\
    \and \hspace{-1.4 cm} Ronald J. Bosch, \\ \vspace{0.cm}
\hspace{-2cm} Center for Biostatistics in AIDS Research, Harvard T.H. Chan School of Public Health \\ 
\and Judith J. Lok \\
    Department of Mathematics and Statistics, Boston University\\
     \vspace{-1.0cm} }

\date{\today}

 \maketitle

\medskip
\vspace{-0.5cm} 

\begin{abstract}
Many clinical studies evaluate the benefit of a treatment based on both survival and other continuous/ordinal clinical outcomes, such as Quality of Life scores. In these studies, when subjects die before the follow-up assessment, the clinical outcomes become undefined and are truncated by death. Treating outcomes as “missing” or “censored” due to death can be misleading for treatment effect evaluation. We show that if we use the median in the survivors or in the always-survivors as estimands to summarize clinical outcomes, we may conclude that a trade-off exists between the probability of survival and good clinical outcomes, even in settings where both the probability of survival and the probability of any good clinical outcome are better for one treatment. Therefore, we advocate not always treating death as a mechanism through which clinical outcomes are missing, but rather as part of the outcome measure. To account for the survival status, we describe the survival-incorporated median as an alternative summary measure for outcomes in the presence of death. The survival-incorporated median is the threshold such that 50\% of the population is alive with an outcome above that threshold. Through conceptual examples and an application to a prostate cancer treatment study, we show that the survival-incorporated median provides a simple and useful summary measure to inform clinical practice.

\end{abstract}

\vfill

\newpage
\spacingset{1.45} 
\section{Introduction}

In clinical practice, treatment decisions are often based on both the probability of survival and other clinical outcomes, such as Quality of Life (QoL) scores or cognitive outcomes. For example, regarding the clinical benefit of a treatment in patients with severe prostate cancer, treatment decisions depend on (1) whether the treatment improves survival and (2) whether the treatment improves Quality of Life \citep{petrylak2004docetaxel, shrestha2019quality}. When interest in a research study lies in both survival and other clinical outcomes, death before the follow-up assessment results in undefined clinical outcomes. In the prostate cancer example, if a patient dies before the follow-up assessment of Quality of Life, his QoL score is undefined. Many authors refer to this setting as “truncation by death” \citep{mcconnell2008truncation,imai2008sharp,kurland2009longitudinal,chiba2011simple} to distinguish it from settings where the outcome is simply missing. 

It is well-known that analyses without careful consideration of undefined outcomes due to death could lead to flawed treatment decisions \citep{robins1995analytic, frangakis1999addressing, zhang2003estimation}. For clinical outcomes in the presence of death, many studies simply regard subjects who die as missing and exclude them from the analyses. Such analyses, where the comparison of treatment options is restricted only to the survivors, will not lead to causally interpretable results, since the comparison of outcomes in survivors is made between two subpopulations that are inherently different \citep{wang2017identification}: those who survive under one treatment versus those who survive under the other treatment.

To appropriately account for truncation by death, \cite{robins1986new} proposed estimating the average causal effect in the always-survivors: the subpopulation that would have survived regardless of their treatment option. The idea was later formalized by \cite{frangakis1999addressing, frangakis2002principal} as principal stratification, which classifies subjects into different subgroups by their potential values of a post-treatment variable. When the post-treatment variable is death, subjects can be classified into four subgroups: always-survivors, protected, harmed, and never-survivors. The survivor average causal effect (SACE) is defined as the treatment effect on the clinical outcomes in the always-survivors. Since the clinical outcomes of the always-survivors are well-defined under both treatment options, the SACE is causally interpretable in the presence of death. However, Section \ref{sec4} shows that the SACE can be a misleading summary measure when evaluating the clinical benefits of treatments.

In this article, we advocate not always treating death as a mechanism through which clinical outcomes are missing or censored, but rather as part of the outcome measure. We propose summarizing the clinical benefit of a treatment by combining death and the clinical outcome into a composite outcome \citep{lachin1999worst, joshua2005treatment, wang2017inference}. To create the composite outcome, we rank all outcomes. In the prostate cancer example, we consider the following ranking: (1) subjects who die are considered to have a worse outcome than survivors with any QoL score, and (2) in survivors, lower QoL scores are considered worse than higher QoL scores. Because the ranked outcomes are a composition of two types of outcomes with two different scales (death and clinical outcomes), instead of using means, it is more appropriate to draw inference by comparing the distribution or quantiles of the composite outcome. Therefore, we propose to use the median of the composite outcome, the survival-incorporated median \citep{lok2010long}, to inform clinical practice.

In the presence of death, the composite outcome approach is also highlighted by the International Committee for Harmonization (ICH) of Technical Requirements for Pharmaceuticals for Human Use, in \citet{ich2017estimands}. This recent guideline on statistical principles for clinical trials indicates: “Terminal events, such as death, are perhaps the most salient examples of the need for the composite strategy.” In the setting of truncation by death, ICH E9(R1) recommends using composite outcomes to define the estimand, since the clinical question of interest is survival along with clinically relevant outcomes.

Researchers have proposed quality-adjusted life year (QALY) \citep{weinstein1973critical, weinstein2009qalys} to measure the quality and the quantity of life lived. QALY year is usually analyzed in means instead of medians. Compared to QALY, the survival-incorporated median does not need to define the state of perfect health, and can be applied not only to QoL scores but also to other clinical outcomes. 

\citet{colantuoni2018statistical} have discussed and compared survivors-restricted analyses, the SACE, and composite outcome approaches using a randomized controlled trial \citep{girard2008efficacy} of critically ill patients with outcomes truncated by death. They concluded that the three statistical approaches can lead to different conclusions if the treatment affects mortality, hence they suggest that careful consideration is needed when analyzing and interpreting the results of clinical studies with mortality. Despite the importance of choosing the optimal approach when analyzing data with truncation by death, in the current literature there is still minimal description of settings where survivors-restricted analyses or SACE can lead to misleading guidance for clinical practice. Sections \ref{sec3}-\ref{sec5} discuss such settings, and we show that in such settings, the survival-incorporated median can be especially useful as a summary measure. 

This article is organized as follows. Section \ref{sec2} defines the survival-incorporated median. Section \ref{sec3} compares the survival-incorporated median with the median in the survivors using an illustrative example. Section \ref{sec4} compares the survival-incorporated median with the median in the always-survivors using the same illustrative example. Section \ref{sec5} explains that the usefulness of the different clinical outcome measures depends on the direction of the effects. Section \ref{sec6} illustrates the survival-incorporated median through simulation studies. Section \ref{sec7} applies the survival-incorporated median to compare the clinical benefits of two treatments for prostate cancer patients using the Southwest Oncology Group (SWOG) S9916 study \citep{petrylak2004docetaxel}. A discussion concludes this article.

\section{The survival-incorporated median: definition}\label{sec2}

The survival-incorporated median is a summary measure of the ranked composite outcome that combines death and ordinal/continuous clinical outcomes such as the QoL score 12 months after initiation of cancer treatment, or neurocognitive scores measured during the study. The composite outcome strategy has been proposed and discussed in previous research \citep{lachin1999worst, joshua2005treatment, wang2017inference}. \citet{dream2006effect} have calculated medians by assigning people with diabetes the worst rank score to study the effect of Ramipril on the incidence of diabetes. \citet{lok2010long} used the survival-incorporated median to study the effect of HIV treatment on CD4+ T-cell counts. In settings where the probability of survival is greater than 50\%, the survival-incorporated median \citep{lok2010long} is defined as:
\begin{definition}[survival-incorporated median]
    \text{} The threshold such that 50\% of the target population is alive with an outcome above that threshold, and 50\% is either dead or has a worse outcome.
\end{definition}

To calculate the survival-incorporated median, we rank all outcomes. The outcome ranking may depend on the clinical context. Typically, subjects who die will be ranked lower than survivors. Among survivors, unfavorable clinical outcomes are ranked lower than favorable clinical outcomes \citep{felker2010global}. To facilitate computation, we could assign those who die a value less than the worst clinical outcome among survivors. For example, if the clinical outcomes are positively valued measurements, we could assign -1 to subjects who die. Combining death and clinical outcomes together, the survival-incorporated median is simply the 50th quantile of the ranked composite outcome.

\begin{figure}
    \centering
    \includegraphics[width = 0.95\textwidth]{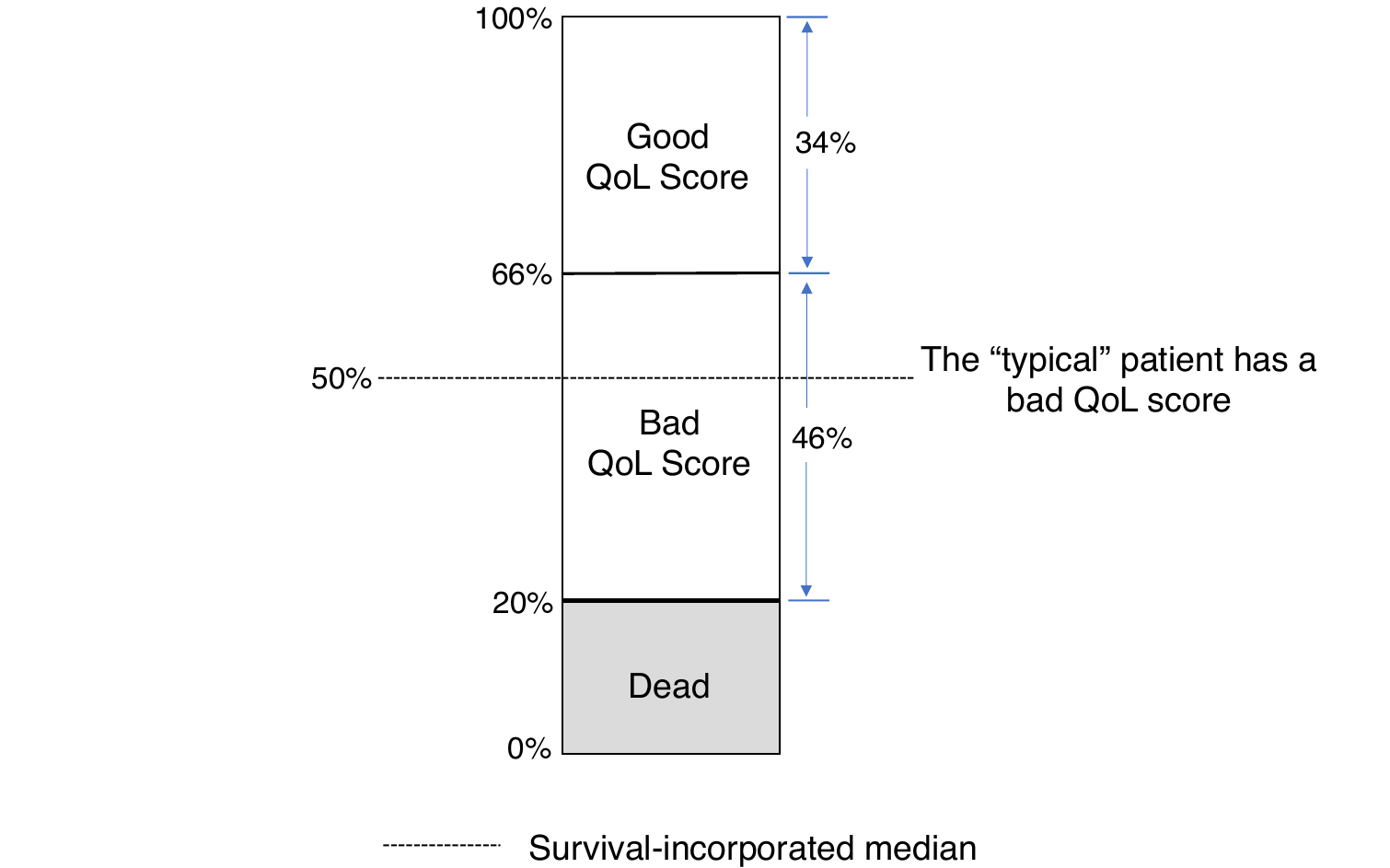}
    \caption{The survival-incorporated median. QoL score: Quality of Life score.}
    \label{fig:chap2 illustration of SIM}
\end{figure}

Figure \ref{fig:chap2 illustration of SIM} illustrates the ranked composite outcome that determines the survival-incorporated median. In Figure \ref{fig:chap2 illustration of SIM}, the probability of death is 20\%, the probability of survival with a bad QoL score is 46\%, and the probability of survival with a good QoL score is 34\%. To create the ranked composite outcome, first, we rank subjects who die lower than subjects who survive. Then, we rank the survivors with bad QoL scores lower than the survivors with good QoL scores. Here, the survival-incorporated median – the 50th quantile of the ranked composite outcome – is a bad QoL score. It represents the status of the “typical/median” subject.

The survival-incorporated median can be interpreted as what happens to the ``typical" subject. For example, if fewer than 50\% of the subjects die, 50\% of the subjects survive with a clinical outcome better than the survival-incorporated median, and 50\% of the subjects die or have a clinical outcome worse than the survival-incorporated median. If more than 50\% of the subjects die, the median of the ranked composite outcome is “death,” since the “typical” subject dies. In such settings, we can consider other survival-incorporated quantiles. For example, if the probability of death is 60\%, the survival-incorporated 75th or 90th quantile could be used as a summary measure of the clinical outcome.

\begin{figure}
    \centering
    \includegraphics[width = 0.7 \textwidth]{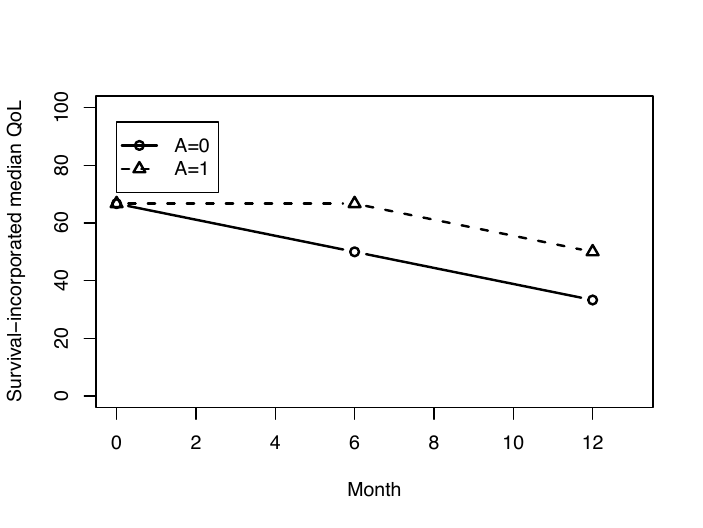}
    \caption{The survival-incorporated median QoL score estimated at baseline, month 6, and month 12 in SWOG study. QoL: Quality of Life score. SWOG: Southwest Oncology Group. A=0: mitoxantrone and prednisone. A=1: docetaxel and estramustine.}

\label{fig:2.6}
\end{figure}

When a study has multiple time points where the clinical outcome is measured, the survival-incorporated median can be either estimated at the end of the study or at any time point(s) during the study, as long as more than 50\% of participants are alive at the time(s) of interest. For each time point, the survival-incorporated median describes the clinical outcome incorporating deaths prior to that time point. See Figure \ref{fig:2.6} for the survival-incorporated median QoL scores estimated at baseline, month 6, and month 12 using SWOG data.

\section{The survival-incorporated median versus the median in the survivors }\label{sec3}

\begin{figure}
    \centering
    \includegraphics[width = 1.05\textwidth]{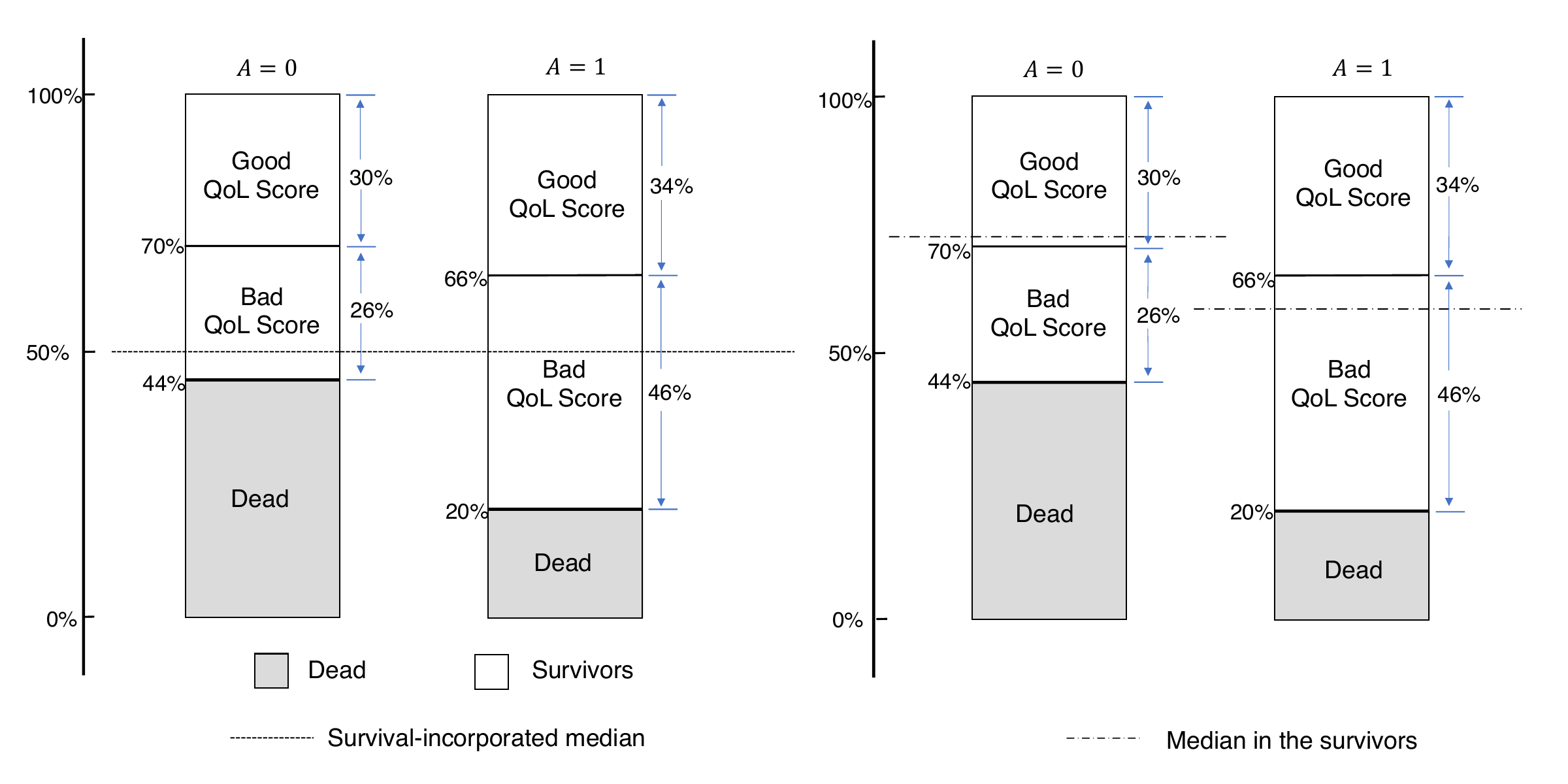}
    \caption{The survival-incorporated median (left) versus the median in the survivors (right). QoL score: Quality of Life score.}
    \label{fig:2.2}
\end{figure}

Figure \ref{fig:2.2} describes an example to compare the survival-incorporated median (left panel) and the median in the survivors (right panel) in the same clinical setting. Suppose we are interested in the effect of treatment in a clinical study with two treatment arms, where the clinical outcomes can be classified as: “good QoL score” versus “bad QoL score.” In Figure \ref{fig:2.2}, for treatment $A=0$ versus $A=1$, the probability of death is 44\% versus 20\%, the probability of survival with a bad QoL score is 26\% versus 46\%, and the probability of survival with a good QoL score is 30\% versus 34\%.

In Figure \ref{fig:2.2}, the median in the survivors (Figure \ref{fig:2.2}, right) is conditional on a post-treatment outcome, survival. For $A=0$ it is a good QoL score, while for $A=1$ it is a bad QoL score. The probability of death is higher for $A=0$ than for $A=1$. Thus, the effect of treatment on Quality of Life in the survivors is in the opposite direction to the effect of treatment on survival. Contrasting the probability of death and the median QoL score in the survivors, we would conclude that there seems to be a trade-off between survival and QoL scores when deciding between $A=0$ versus $A=1$.

We reach a different conclusion based on the survival-incorporated median (Figure \ref{fig:2.2}, left). In the ranked composite outcome, the threshold that separates good QoL scores from bad QoL scores is the 70th quantile for $A=0$ and the 66th quantile for $A=1$. The survival-incorporated median, the 50th quantile in this ranking, is a bad QoL score under both treatments. In contrast to the median in the survivors, the survival-incorporated median does not suggest a trade-off between survival and QoL scores. Moreover, compared with the survival-incorporated median for $A=0$, the survival-incorporated median for $A=1$ is closer to the threshold which separates good and bad QoL scores, suggesting subjects may have better QoL scores under $A=1$.  

What can we conclude? From Figure \ref{fig:2.2}, the probability of survival with a good QoL score is 30\% for $A=0$ versus 34\% for $A=1$. The probability of death is 44\% for $A=0$ versus 20\% for $A=1$. Thus, both the probability of survival and the probability of survival with a good QoL score are better for $A=1$ than for $A=0$. While the median in the survivors failed to conclude this, the survival-incorporated median does not point in the wrong direction. In fact, the survival-incorporated 68th quantile is higher under $A=1$ than under $A=0$, reflecting the slightly higher probability of surviving with a good QoL score under $A=1$. 

\section{The survival-incorporated median versus the median in the always-survivors}\label{sec4}

\subsection{Review of the survivor average causal effect (SACE)}

The survivor average causal effect (SACE), also called the principal strata effect, provides a summary measure of clinical outcomes truncated by death \citep{robins1986new, frangakis2002principal}. To illustrate principal stratification, let $S(0)$ and $S(1)$ be indicators of the potential survival status of a subject if the subject received treatment $A=0$ and $A=1$, respectively. For example, $S(1)=1$ if the subject would have survived under $A=1$, and $S(1)=0$ if not. Let $G$ denotes the principal stratum that a subject belongs to. With principal stratification, subjects are categorized into four subgroups (Table \ref{tab:2.1}):

\begin{table}[h]
  \centering
  	\renewcommand{\arraystretch}{1.4}
  \begin{tabular}{ p{0.65cm} p{0.65cm} p{3cm} p{11 cm}  } 
  \hline
    \textbf{$S(0)$} & \textbf{$S(1)$} & \textbf{Survival type} & \textbf{Description} \\ \hline
    1 & 1 & Always-survivor & The subject always survives regardless of treatment. \\ 
    0 & 1 & Protected & The subject dies under control, but survives under treatment. \\ 
    1 & 0 & Harmed & The subject survives under control, but dies under treatment. \\ 
    0 & 0 & Never-survivor & The subject never survives regardless of treatment. \\ 
    \hline
  \end{tabular}
    \caption{Four principal strata in the presence of death.}

\label{tab:2.1}
\end{table}

The SACE focuses on assessing the benefit of a treatment for a subgroup of subjects: the always-survivors. However, the always-survivors often do not represent the entire target population, and who would be an always-survivor is usually not known at the time the treatment decisions are made. Hence, the treatment effect in this subgroup is not identifiable without further assumptions. In various studies of identification and estimation of the SACE \citep{chiba2011simple, wang2017identification, gilbert2003sensitivity, hayden2005estimator, egleston2007causal, shepherd2008does, ding2011identifiability, tchetgen2014identification} assumptions usually include the principal ignorability assumption \citep{stuart2015assessing} and the monotonicity assumption:

\begin{assumption}[principal ignorability assumption]
     $P(G=g|L,Y(0),Y(1) )=P(G=g|L). $
\end{assumption}

\begin{assumption}[monotonicity]
    For all individuals, $S(0) \leq S(1)$.
\end{assumption}

The principal ignorability assumption states that principal stratum membership $G$ is independent of the potential outcomes $(Y(0),Y(1))$ given the observed covariates \citep{stuart2015assessing}. The monotonicity assumption states that, for all individuals, survival under treatment is always at least as good as survival under control. In other words, death under treatment must imply death under control. Hence, under the monotonicity assumption, there are no subjects in the harmed stratum, and there are only three strata of subjects: always-survivors, protected, and never-survivors.

\subsection{The survival-incorporated median versus the median in the always-survivors }

\begin{figure}
    \centering
    \includegraphics[width = 1.05 \textwidth]{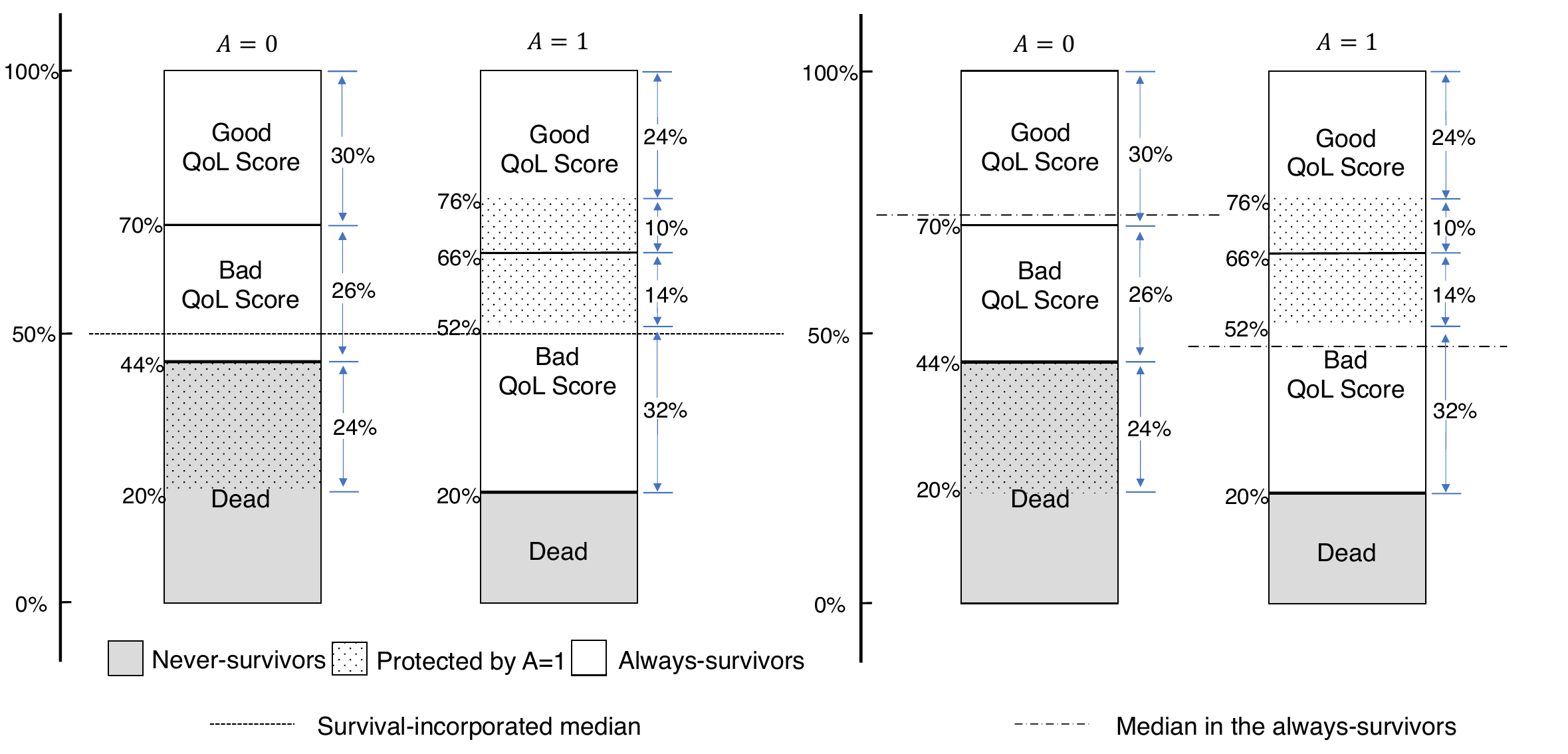}
    \caption{The survival-incorporated median (left) versus the median in the survivors (right). QoL score: Quality of Life score.}
    \label{fig:2.3}
\end{figure}

Is it possible that the median in the always-survivors could be misleading to inform treatment decisions aimed at clinical practice? Figure \ref{fig:2.3} describes such an example under the monotonicity assumption. In Figure \ref{fig:2.3}, the probabilities of death and the probabilities of survival with a good QoL score are the same as in Figure \ref{fig:2.2}. The difference is that we now mark the stratum that a subject belongs to. Because of monotonicity, under $A=0$, all survivors are the always-survivors, while under $A=1$, survivors include both the always-survivors and the protected.  The protected in Figure \ref{fig:2.3} are the 24\% of subjects who would die under $A=0$ but survive under $A=1$. In Figure \ref{fig:2.3}, under $A=1$, 10\% of subjects are protected with a good QoL score and 14\% of subjects are protected with a bad QoL score; 24\% of subjects are always-survivors with a good QoL score and 32\% of subjects are always-survivors with a bad QoL score.

In Figure \ref{fig:2.3}, the median in the always-survivors (right panel) for $A=0$ is a good QoL score, while for $A=1$ it is a bad QoL score. The probability of death is higher for $A=0$ than for $A=1$. Thus, the effect of treatment on Quality of Life in the always-survivors is in the opposite direction to the effect of treatment on survival. Similar to the median in the survivors (Section \ref{sec3}), contrasting the probability of death and the median QoL score in the always-survivors, we would conclude that there seems to be a trade-off between survival and QoL scores when deciding between $A=0$ versus $A=1$.

As in Section \ref{sec3}, the survival-incorporated median (Figure \ref{fig:2.3}, left) is a bad QoL score under both treatments. In contrast to the median in the always-survivors, the survival-incorporated median does not suggest a trade-off between survival and good QoL scores. In Figure \ref{fig:2.3}, both the probability of survival and the probability of survival with a good QoL score are better for $A=1$ than for $A=0$. The median in the always-survivors points in the wrong direction, but the survival-incorporated median does not. See also the discussion at the end of Section \ref{sec3}.

\section{Direction of the effects} \label{sec5}

Clinical decisions are often based on both survival and clinical outcomes. We argue that the usefulness of the clinical outcome measures in the survivors/always-survivors depends on whether the effect on survival is in the same direction as the effect on the clinical outcome measures in the survivors/always-survivors. In some cases, optimal treatment decisions for the always-survivors may not be optimal for all patients. In particular, in settings where the treatment effect on survival is in the opposite direction of the treatment effect on the clinical outcome measures in the survivors/always-survivors, we should be cautious about using those measures. Sections \ref{sec3} and \ref{sec4} describe a setting where both the probability of survival and the probability of survival with a good QoL score are better for $A=1$, but the median clinical outcome in the survivors/always-survivors is worse for $A=1$. In such settings, we may falsely conclude that there is a trade-off between survival and clinical outcomes. However, the survival-incorporated median, which may or may not suggest such a trade-off, can be used to summarize whether overall the clinical outcomes are better under a particular treatment.

  \begin{figure}
    \centering
    \includegraphics[width = 1.05 \textwidth]{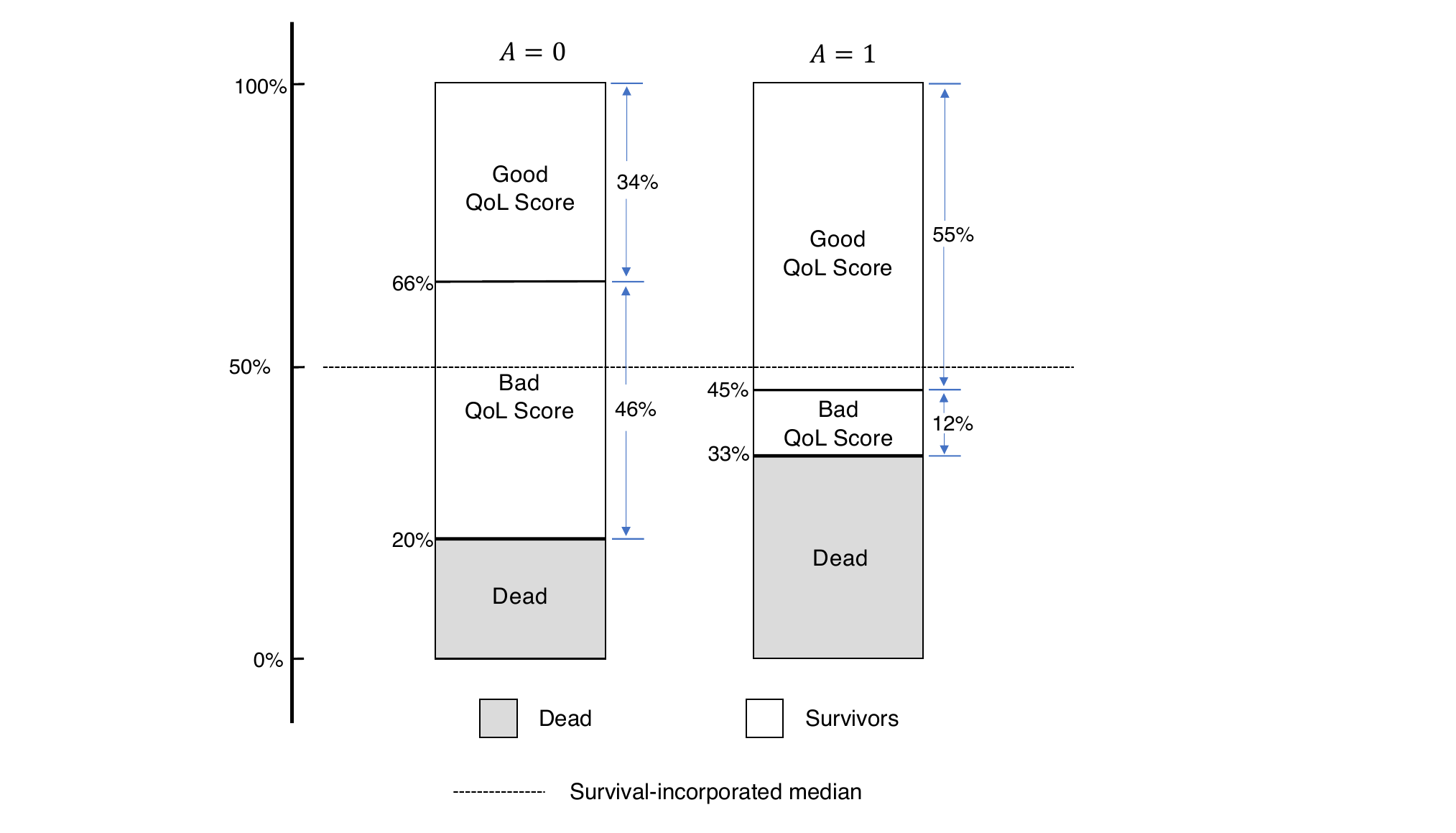}
    \caption{Setting where the treatment effect on survival is in the opposite direction of the treatment effect on the survival-incorporated median}
    \label{fig:2.4}
\end{figure}

In some settings, the effect of treatment on survival is not in the same direction as the effect of treatment based on the survival-incorporated medians. Also in such settings, the survival-incorporated approach can inform clinical practice. For example, in Figure \ref{fig:2.4}, the probability of death is higher under $A=1$, but $P(alive\ with\ good\ score|A=1)$ is greater than $P(alive\ with\ good\ score|A=0)$. The survival-incorporated median is a good score for $A=1$ and it is a bad score for $A=0$. Thus, in this setting, the survival-incorporated medians reveal that the treatment effect on survival is indeed in a different direction from the treatment effect on the clinical outcome. In this setting, combining analyses of the survival-incorporated median and the probability of death, one’s decision should depend on whether to optimize towards QoL scores or towards survival.
In practice, we will often not be certain of the directions of the treatment effects on survival and clinical outcomes, especially when at least one of the confidence intervals (for the difference between the survival probabilities or for the difference between the clinical outcome measures) includes zero. Regardless of the direction of effects, the survival-incorporated median can be used to inform clinical practice.

\section{Simulation study} \label{sec6}

\begin{figure}
    \centering
    \includegraphics[width = 0.8\textwidth]{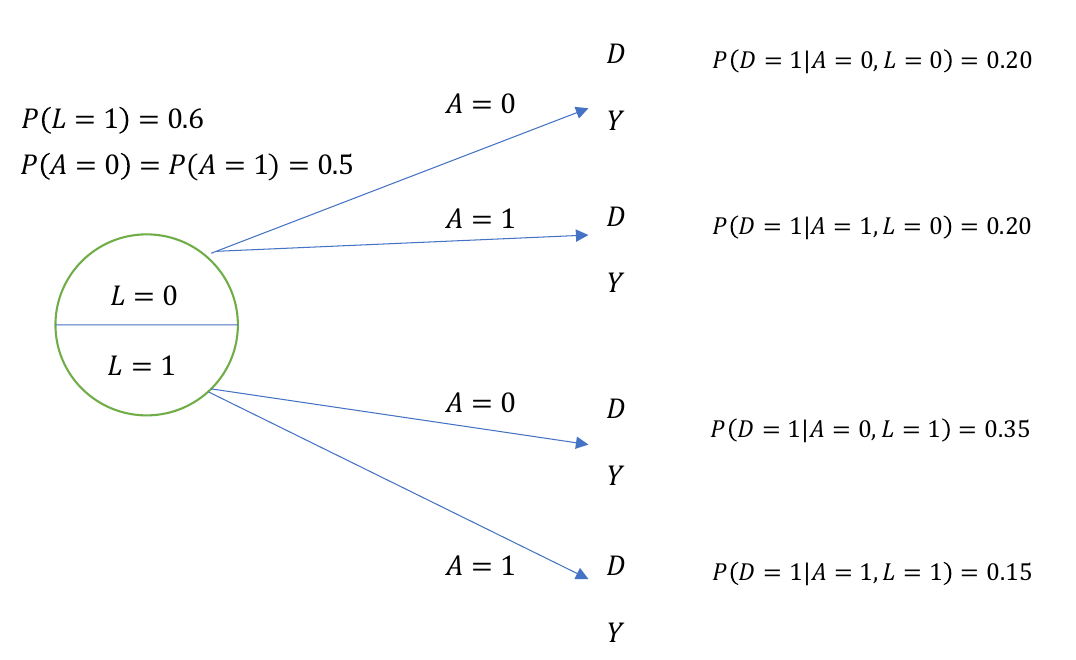}
    \caption{Simulation setting for a randomized trial.}
    \label{fig:2.5}
\end{figure}

We simulate a randomized trial to illustrate the survival-incorporated median. After assigning $A$ randomly, we observe death $(D)$ or a continuous clinical outcome $Y$ (Figure \ref{fig:2.5}). In this simulated setting, treatment $A = 1$ improves survival in patients with $L=1$ but does not affect survival in patients with $L=0$. Assuming that higher values of $Y$ are better than lower values of $Y$, and the clinical outcome $Y$ in the survivors is better under $A=1$; we simulate the clinical outcome in those alive by
\begin{align*}
Y = \beta_0 + \beta_1 \cdot A + \beta_2 \cdot L + \varepsilon,
\end{align*}
where $\beta_0 = 3$, $\beta_1 = 0.3$, $\beta_2 = -3$, and $\varepsilon \sim N(0, 1)$.

For this simulated setting, we mathematically derive the true population survival-incorporated median and the median in the survivors under both $A = 0$ and $A = 1$, see Appendix Section \ref{A.sec2}. The true population survival-incorporated median under $A = 0$ is 0.093 and under $A = 1$ it is 0.670. Thus, treatment $A=1$ is better for the “typical/median” patient. The true population median in the survivors under $A = 0$ is 1.184 and under $A = 1$ it is 1.155, which is misleading because it conditions on the post-treatment survival outcome.

\begin{table}[h] \label{tab:simulation-results}
\centering

\begin{tabular}{|cc|ccc ccc|}
\hline
        &        & \multicolumn{3}{c}{Survival-incorporated median} & \multicolumn{3}{c|}{Median in the survivors} \\
        &        & Truth  & rMSE  & Bias   & Truth  & rMSE  & Bias \\
\hline
$A=0$        & N=500  & 0.093  & 0.207 & 0.001  & 1.184  & 0.261 & 0.010 \\
             & N=1500 &        & 0.116 & -0.001 &        & 0.148 & 0.007 \\
             & N=5000 &        & 0.063 & 1e-03  &        & 0.082 & 9e-04 \\
$A=1$        & N=500  & 0.670  & 0.165 & 0.007  & 1.155  & 0.190 & 0.009 \\
             & N=1500 &        & 0.094 & 0.006  &        & 0.110 & 0.005 \\
             & N=5000 &        & 0.052 & 9e-04  &        & 0.060 & 6e-04 \\
$A=1$-$A=0$  & N=500  & 0.577  & 0.262 & 0.006  & -0.029 & 0.317 & -0.001 \\
             & N=1500 &        & 0.150 & 0.006  &        & 0.182 & -0.003 \\
             & N=5000 &        & 0.082 & -1e-04 &        & 0.102 & -3e-04 \\
\hline
\end{tabular}
\caption{Simulation results for the survival-incorporated median versus the median in the survivors in the clinical trial setting of Section \ref{sec6}. rMSE: root Mean Square Error.}
  \label{tab:2.2}
\end{table}

Table \ref{tab:2.2} summarizes the results of the simulation study for the survival-incorporated median versus the median in the survivors. In each simulation, deaths were assigned a value lower than any observed $Y$. We used the “median” function in R to obtain the estimated results. As expected, Table \ref{tab:2.2} shows that when the sample size is increasing, both the rMSE and bias decrease.

\section{Application: comparing two treatments affecting Quality of Life of prostate cancer patients}\label{sec7}

We analyzed the Southwest Oncology Group (SWOG) S9916 data to illustrate the survival incorporated median. This randomized phase III trial compared two treatment regimens for metastatic, hormone-independent prostate cancer: mitoxantrone and prednisone (MP) and docetaxel and estramustine (DE) \citep{petrylak2004docetaxel}. A total of 674 eligible subjects were randomly assigned to MP or DE. The primary outcome was the survival time, and the secondary outcome was the Quality of Life (QoL) score. In our analysis, we are interested in comparing the clinical benefit of MP and DE on the QoL scores at 12 months, and our population of interest is the group of patients for whom it was possible to measure their QoL score at baseline. We incorporated survival and illness, since a substantial number of patients died or were too ill to report a QoL score. Table \ref{tab:2.3} summarizes the number of subjects and the probability of death under each treatment regimen in the preprocessed data.

\begin{table}[h]

\centering
\renewcommand{\arraystretch}{1.3}

\begin{tabular}{c c c}
\hline
& \#subjects & \#death at 12 months (\%) \\
\hline
MP ($A=0$) & 306 & 108 (35.3\%) \\

DE ($A=1$) & 315 & 84 (26.7\%) \\
\hline
\end{tabular}
\caption{Number of subjects and the probability of death under each treatment regimen among subjects whose QoL scores were measured at baseline. MP: treatment of mitoxantrone and prednisone. DE: treatment of docetaxel and estramustine.}

\label{tab:2.3}
\end{table}

QoL scores of some survivors were missing. Hence, we performed data preprocessing steps to handle the missing QoL scores in the survivors. First, 53 subjects whose QoL scores were missing at baseline (those subjects did not have any QoL score measured) were excluded. Then, since QoL scores in this study ranged from 0 to 100, we assigned the QoL score of the subjects who died a value of -10 and the subjects whose score was missing due to illness a value of -5, thus ranking subjects who were too ill higher than those who died but lower than those with any observed QoL score. In the remaining survivors, 102 had missing month-12 QoL scores.

After data preprocessing, we estimated the survival-incorporated median of the month-12 QoL scores, using Inverse Probability of Censoring Weighting (IPCW) \citep{robins1995analytic, robins2000correcting} to account for month-12 QoL scores in survivors that were missing for reasons other than illness. The Appendix provides details. 

\begin{table}[h] 

\centering 
\renewcommand{\arraystretch}{1.3}

\begin{tabular}{l l l}
\hline
\makecell{Estimates \\ (95\% Confidence Interval)} & \makecell{Survival-incorporated median \\ QoL score} & \makecell{Median QoL score \\in the survivors} \\
\hline
MP ($A=0$) & 33.3 [0, 50.0] & 66.7 [50.0, 66.7] \\

DE ($A=1$) & 50.0 [33.3, 50.0]  & 66.7 [58.3, 66.7] \\

Effect difference $A=1$ – $A=0$ & 16.7 [0, 50] & 0 [-8.3, 8.3] \\
\hline
\end{tabular}
\caption{The estimated survival-incorporated median QoL score and the median QoL score in the survivors at 12 months. QoL score: Quality of Life score. MP: treatment of mitoxantrone and prednisone. DE: treatment of docetaxel and estramustine. 95\% Confidence Interval is constructed by non-parametric bootstrap using Efron’s percentile method.}

\label{tab:2.4}
\end{table}

Table \ref{tab:2.4} shows the estimated survival-incorporated median QoL score and the median QoL score in the survivors at 12 months in both treatment arms. The confidence intervals of the estimated results are constructed by non-parametric bootstrap using Efron’s percentile method33. The survival-incorporated median of the month-12 QoL scores is 33.3 for MP and 50.0 for DE. The median month-12 QoL scores in the survivors are the same for MP and DE: 66.7. While the median in the survivors shows no difference between MP and DE, the survival-incorporated median favors the clinical benefit of DE over MP. This is aligned with the original SWOG paper, which reported an overall survival benefit of DE over MP.

We also compared the survival-incorporated median with the Survival Average Causal Effect (SACE). \citet{ding2011identifiability} proposed an estimation method for the SACE and illustrated their method also with the SWOG S9916 data. The data that \citet{ding2011identifiability}  used in their paper are a subset of the original data, including only 487 of the total 674 subjects. In this subset of the data, the probability of death under both treatments is approximately 50\% at 12 months and we used the same data subset for comparison. We estimated the survival-incorporated 75th quantile under both treatments and compared it with the SACE. The survival-incorporated median and the SACE both show a positive but not statistically significant effect of DE over MP, and the effect is larger using the survival-incorporated median (details in Appendix Section \ref{A.sec4}).

\section{Discussion}\label{sec8}

There are many clinical settings where a clinical outcome is truncated or undefined for an individual who dies before the clinical outcome is measured. Often, the clinical interest is survival along with clinical outcomes \citep{ich2017estimands}. In these settings of truncation by death, focusing on summary measures only in the survivors may lead to bias \citep{robins1986new}. To account for subjects’ survival status, we describe the survival-incorporated median as a useful summary measure for outcomes in the presence of death. By combining death and clinical outcomes into a ranked composite outcome, the survival-incorporated median (or quantiles) can be used to inform clinical practice.

The median in the survivors/always-survivors are sometimes misleading as summary measures to inform clinical practice, and the usefulness of the summary measures in the survivors/always-survivors depends on the direction of the effects. In Sections \ref{sec3} and \ref{sec4}, we showed that there are settings where both the probability of survival and the probability of survival with a good clinical outcome are better under $A=1$, but both the median in the survivors and the median in the always-survivors are worse under $A=1$. These settings are further discussed in Section \ref{sec5}: when the effect on survival is in the opposite direction of the effect on the clinical outcome measures in the survivors/always-survivors, we should be cautious about using those measures. In contrast, it is safe to use the survival-incorporated median, which does not depend on the direction of the effects.

The survival-incorporated median is simpler to compute and requires fewer assumptions than the Survivor Average Causal Effect (SACE) and other methods for the effect of treatment in the presence of death \citep{stensrud2020separable}. Aside from the Ignorability and Monotonicity assumptions described in Section \ref{sec4}, identification and estimation of the SACE require additional assumptions. For example, \citet{ding2011identifiability} assumed that in a randomized trial, there is a pre-treatment covariate that (1) has a distribution that is different between the always-survivors and the protected, and (2) does not predict the clinical outcome given principal stratum and treatment group. \citet{tchetgen2014identification} introduced post-treatment covariates that may mediate the treatment effects on survival and clinical outcomes to identify the SACE, and he made an assumption that is a type of cross-worlds assumption \citep{vanderweele2015explanation, lok2021causal}; it relies on two simultaneous but different situations, treatment and no treatment. The required assumptions for identification and estimation of the SACE are often technical and not verifiable in practice.

As the SWOG data application illustrates, the survival-incorporated median is easy to estimate and does not rely on many assumptions, especially when treatment is randomized and in the absence of censoring. When some outcomes are missing, one can apply Inverse Probability of Censoring Weighting to estimate the survival-incorporated median. In addition, in observational data, Inverse Probability of Treatment Weighting \citep{robins2000marginal} can be used to estimate the survival-incorporated median for a causal interpretation \citep{hogan2004marginal, firpo2007efficient}. In contrast, for the SACE, although eliciting expert opinions may help to ensure the assumptions are valid, it is unlikely that the assumptions will be valid in all trials \citep{colantuoni2018statistical}; the monotonicity assumption, for example, may not be valid in the SWOG S9916 study, which compares two active treatments (MP and DE). When monotonicity does not hold, identification and estimation of SACE are more complicated \citep{ding2011identifiability}.

In many settings, the change of clinical outcomes from baseline is of more clinical interest than the clinical outcomes at a certain time point. When using the survival-incorporated median in such settings, to assure that subjects who die are still ranked lowest in the composite outcome, we should assign a value less than the lowest value of the change from baseline. If the clinical outcomes range from 0 to 100, the lowest possible change is -100, so one should assign the month-12 change in the clinical outcomes of a subject who died a value less than -100, for example, -500. In some cases, one should be careful when combining clinical outcomes and mortality, since some severely ill patients may consider their quality of life worse than death. Those patients could be incorporated by assigning them a score less than the score for death. Thus, computing survival-incorporated medians relies on the order of clinical outcomes and mortality, which may require input from both clinicians and patients.

“Truncation by death” also arises in other fields such as economics \citep{lalonde1995promise}, education \citep{zhang2003estimation}, and social science \citep{mcconnell2008truncation}, where the survival-incorporated median can also be a useful summary measure. For example, in an evaluation of a healthy-marriage intervention, the outcome of interest is the quality of marriage index that measures marital satisfaction \citep{mcconnell2008truncation}. When couples are divorced, their quality of marriage index becomes undefined. Since the divorce can be viewed as a truncation of marriage, by combining the marital status and quality of marriage index into a ranked composite outcome, the survival-incorporated median provides a summary measure to evaluate the benefit of the healthy-marriage intervention.

The survival-incorporated median can be applied for planning purposes and also as an analytic approach \citep{lok2010long}. If one foresees that death will happen when planning the trial, then one could plan the trial based on both the probability of death and the survival-incorporated median.

In this chapter, we characterize the survival-incorporated median, which has valuable applications in clinical settings involving longer-term follow-up. We advocate that the survival-incorporated median be broadly used by researchers and practitioners in clinical studies as a useful estimand, summarizing clinical outcomes in the presence of death to inform clinical practice.

\section{Acknowledgement}
This work was sponsored by NSF grant DMS 1854934 to Judith J. Lok and NIH/NIAID grant UM1 AI068634 supporting Ronald J. Bosch. The content is solely the responsibility of the authors and does not necessarily represent the official views of the NSF or the NIH.


\bibliographystyle{abbrvnat}
\bibliography{surMed}

\vskip .65cm

\newpage

\begin{appendices}

\section{The survival-incorporated median versus the median in the always-survivors: example without monotonicity} \label{A.sec1}

\begin{figure}
    \centering
    \includegraphics[width = 1.05 \textwidth]{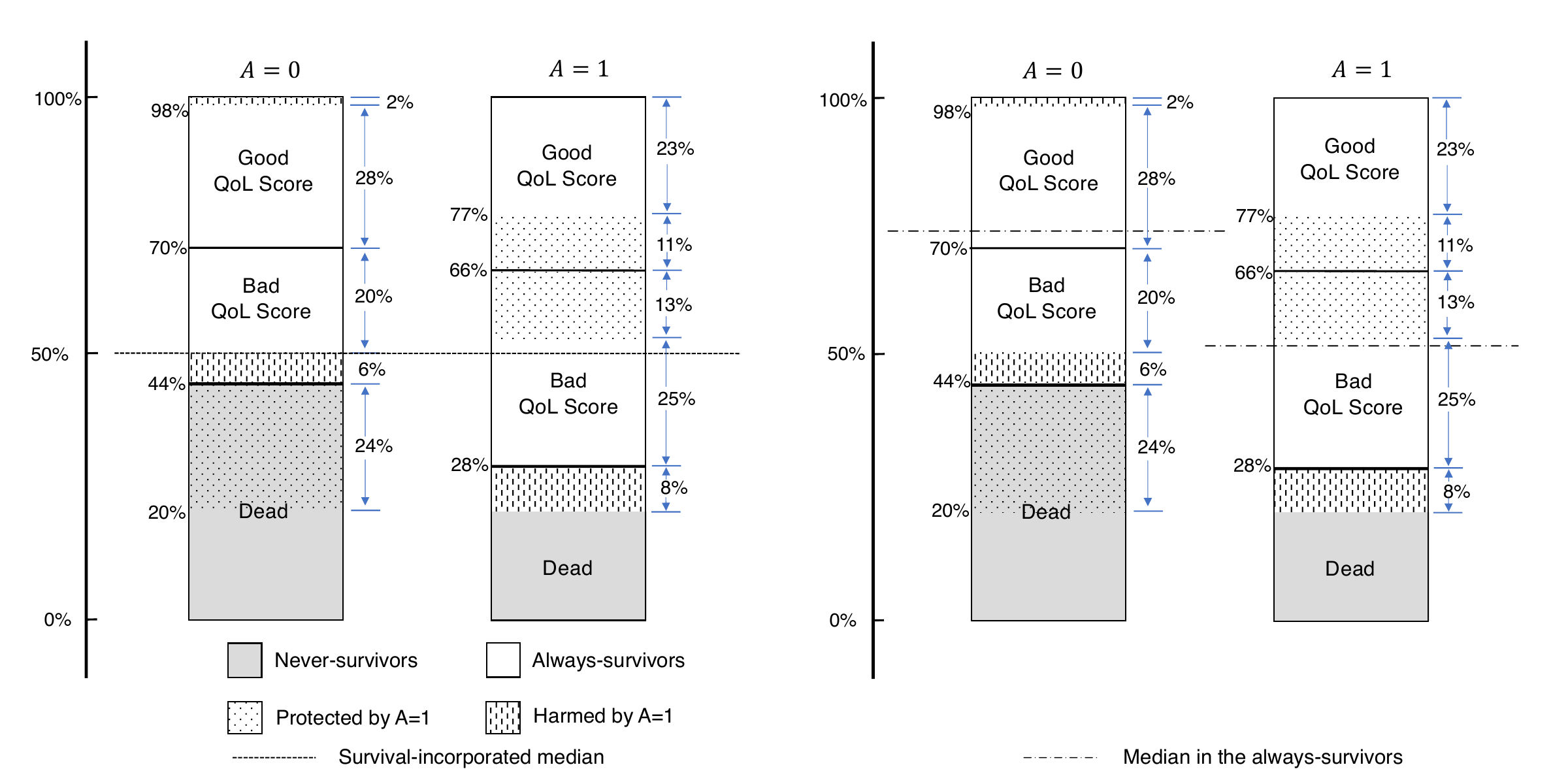}
    \caption{The survival-incorporated median (left) versus the median in the always-survivors (right): example without monotonicity. QoL score: Quality of Life score.}
    \label{fig:A.1}
\end{figure}

Supplementary Figure \ref{fig:A.1} describes a hypothetical example comparing the survival-incorporated median and the median in the always-survivors without the monotonicity assumption. Without monotonicity, all four strata of subjects are present: always-survivors, protected, harmed, and never-survivors. In Supplementary Figure \ref{fig:A.1}, the setting is almost the same as in Figure \ref{fig:2.2} in terms of the probability of death, the probability of a good QoL score in both treatment arms, the percentage of protected subjects, and the percentage of never-survivors. The major difference is that under $A=0$, 2\% of subjects are harmed with a good QoL score and 6\% are harmed with a bad QoL score.

Similar to Section \ref{sec3}, in Supplementary Figure \ref{fig:A.1}, right, the median in the always-survivors for $A=0$ is a good QoL score, while for $A=1$ it is a bad QoL score. Contrasting the probability of death and the median in the always-survivors, we would conclude that there seems to be a trade-off between survival and QoL scores when using the SACE to decide between $A=0$ versus $A=1$. However, the survival-incorporated median (Supplementary Figure \ref{fig:A.1}, left) is a bad QoL score under both treatments. In contrast to the median in the always-survivors, the survival-incorporated median does not suggest a trade-off between survival and good QoL scores. In Supplementary Figure \ref{fig:A.1}, both the probability of survival and the probability of survival with a good QoL score are better for $A=1$ than for $A=0$. This is captured by the survival-incorporated median, not by the median in the always-survivors.

In practice, unlike Supplementary Figure \ref{fig:A.1}, the principal stratum of each subject is unknown, and the assumptions for SACE are difficult to verify using observed data. Without the monotonicity assumption, estimating and identifying SACE requires other assumptions and becomes much more complicated. In contrast, the survival-incorporated median, which does not rely on untestable assumptions, is easy to compute and often leads to stable estimates.

\section{Derivation of the truth in the simulation studies} \label{A.sec2}

We derive the truth mathematically for the simulation setting in Figure \ref{fig:2.5}. For the distribution of the composite outcome under treatment $A = 0$,

\begin{align*}
& F_{Y^{(0)}}(y) \\
& = \sum_{l\in(0,1)} P(L=l) \sum_{d\in(0,1)} P(D^{(0)}=d|L=l) \cdot F_{Y^{(0)}|D^{(0)}=d,L=l} (y) \\
& = P(L=0) \cdot P(D^{(0)}=1|L=0) \cdot 1 + P(L=1) \cdot P(D^{(0)}=1|L=1) \cdot 1 + \\
& \qquad P(L=0) \cdot P(D^{(0)}=0|L=0) \cdot F_{Y^{(0)}|D^{(0)}=0,L=0(y)} + \\
& \qquad P(L=1) \cdot P(D^{(0)}=0|L=1) \cdot F_{Y^{(0)}|D^{(0)}=0,L=1}(y) \\
& = 0.4 \cdot 0.2 \cdot 1 + 0.6 \cdot 0.35 \cdot 1 + \\  
& \qquad  0.4 \cdot 0.8 \cdot F_{Y^{(0)}|D^{(0)}=0,L=0}(y) + 0.6 \cdot 0.65 \cdot F_{Y^{(0)}|D^{(0)}=0,L=1}(y) \\
& = 0.29 + 0.32F_{Y^{(0)}|D^{(0)}=0,L=0} + 0.39F_{Y^{(0)}|D^{(1)}=0,L=1},
\end{align*}
where $F_{Y^{(0)}|D^{(0)} = 0, L = 0} \sim N(3,1)$ and $F_{Y^{(0)}|D^{(0)} = 0, L = 1} \sim N(0,1)$ are the CDFs of the normal distributions for those alive under $A = 0$ with $L = 0$ and $L = 1$, respectively.

Accordingly, the survival-incorporated median under $A = 0$ is $F_{Y^{(0)}}^{-1}(0.5)$.

Similarly, under treatment $A = 1$,
\begin{align*}
    F_{Y^{(1)}}(y) = 0.17 + 0.32F_{Y^{(1)}|D^{(1)} = 0, L = 0} + 0.51F_{Y^{(1)}|D^{(1)} = 0, L = 1},
\end{align*}
where $F_{Y^{(1)}|D^{(1)} = 0, L = 0} \sim N(3.3,1)$ and $F_{Y^{(1)}|D^{(1)} = 0, L = 1} \sim N(0.3,1)$ are the CDFs of the normal distributions for those alive under $A = 1$ with $L = 0$ and $L = 1$, respectively.

The survival-incorporated median under $A = 1$ is 
$$
F_{Y^{(1)}}^{-1}(0.5).
$$

Deriving the median in the survivors is similar. The difference is that now the CDF of interest is conditional on $D = 0$:
\begin{align*}
   F_{Y^{(0)}|D^{(0)} = 0}(y) = \frac{32}{71}F_{Y^{(0)}|D^{(0)} = 0, L = 0} + \frac{39}{71}F_{Y^{(0)}|D^{(0)} = 0, L = 1} \\
   F_{Y^{(1)}|D^{(1)} = 0}(y) = \frac{32}{83}F_{Y^{(1)}|D^{(1)} = 0, L = 0} + \frac{51}{83}F_{Y^{(1)}|D^{(1)} = 0, L = 1}.
\end{align*}

The median in the survivors under $A = 0$ and $A = 1$ are $F_{Y^{(0)}|D^{(0)} = 0}^{-1}(0.5)$ and $F_{Y^{(1)}|D^{(1)} = 0}^{-1}(0.5)$, respectively.

\section{Data analysis details} \label{A.sec3}
\subsection{Data preprocesing}

The data preprocessing steps for the SWOG data include (1) excluding subjects who could not or did not report their Quality of Life (QoL) score at baseline, and (2) how we deal with missing clinical outcomes in survivors. We received a dataset with 650 out of the original 674 subjects, where 24 subjects were already excluded due to the missing QoL score at baseline. The data, which have a complete follow-up of survival until 395 days, include QoL scores (ranging from 0 to 100) at baseline, month-3, month-6, and month-12. In these 650 subjects, to compute the survival-incorporated median, we assigned negative scores (any number below the lowest QoL score can be used here) to subjects who died or had missing QoL score due to illness. Some of the survivors have missing QoL scores for different reasons: institution error, refused phone call, and other reasons. To account for missing values in survivors, we preprocessed the data in the following steps:

\begin{enumerate}
    \item We excluded 29 subjects whose QoL score is missing at baseline; actually, those subjects do not have QoL scores measured at any follow-up time.
    \item If a subject is alive at 365 days and has a month-12 QoL score, we used it.
    \item If a subject is dead at 365 days, their status is “death” (we assigned QoL = -10). 
    \item If a subject is alive at 365 days with missing month-12 QoL score due to illness, their status is “illness” (we assigned QoL = $-5$).
    \item If a subject is alive at 365 days with a missing month-12 score but is dead before 395 days, we regarded those subjects as missing due to illness (we assigned QoL = -5).
    \item If a subject is alive at 395 days with a missing month-12 score, but the subject’s month-6 score is not missing, we used the month-6 score as their month-12 score.
    \item There are 52 survivors with missing QoL scores due to institution error, which are true missings, and we assumed Missing Completely At Random (MCAR) \citep{rubin1976inference}. 
    \item  There are still 50 survivors with missing QoL scores (other reasons, refused phone call, or no reason recorded), and we assumed Missing At Random (MAR) \citep{rubin1976inference}. 
    
\end{enumerate}

After data preprocessing, each subject has a valid QoL score, or an assigned QoL score incorporating death and illness, or a missing QoL score that we accounted for. Subjects who died have an assigned month-12 QoL scores of -10, and subjects with the missing month-12 QoL score due to illness have an assigned score of -5. Most survivors have a valid month-12 QoL score, and for survivors with missing month-12 QoL scores, we used the Inverse Probability of Censoring Weighting (IPCW) \citep{robins1995analytic, robins2000marginal} to account for their missing values.

\subsection{Inverse Probability of Censoring Weighting (IPCW) for missing Quality of Life scores}

We now show the details of the choice of weights for every subject when applying IPCW, so that the weighted population resembles the overall population. We first introduce some notation. $i$ is the index of $i$th subject, where $i=0,…,621$. $L_i$ are the baseline covariates of subject $i$: age, race, performance status, and baseline QoL score. $A_i$ is an indicator for subject $i$’s treatment, where $A_i=0$ represents the MP arm and $A_i = 1$ represents the DE arm. In the application in the main text, we consider the illness-and-survival incorporated QoL score. $D_{(i,death)}$ is an indicator of the survival status of subject $i$, where $D_{(i,death)}=1$ if subject $i$ died before month-12 and $D_{(i,death)}=0$ if not. $D_{(i,ill)}$ is an indicator of the illness status of subject $i$, where $D_{(i,ill)}=1$ if a subject’s QoL score is unmeasured due to illness and $D_{(i,ill)}=0$ if not. $C_{(i,insti)}$ is an indicator for missing due to institution error. $C_{(i,other)}$ is an indicator for missing due to refused phone calls, other reasons, or no reason recorded.

In those alive, we assumed Missing Completely At Random (MCAR) for $C_{(i,insti)}$ and Missing At Random (MAR) for $C_{(i,other)}$. IPCW is used to account for all subjects who initiated the treatments with non-missing baseline QoL scores. IPCW proceeds as follows:

\begin{enumerate}
    \item The censoring probability for survivors with missing values due to institution error is, since we assumed MCAR in those alive,
\begin{align*}
& p(C_{(i,insti)}=1|L_i,A_i,D_{(i,death)}=0,D_{(i,ill)}=0) \\
 = & p(C_{(i,insti)}=1|A_i,D_{(i,death)}=0,D_{(i,ill)}=0). 
\end{align*}

    We estimated this probability under each treatment $a$ by
\[
\frac{\#\{C_{(i,insti)}=1,A_{i}=a,D_{(i,death)}=0,D_{(i,ill)}=0\}}{\#\{A_{i}=a,D_{(i,death)}=0,D_{(i,ill)}=0\}}.
\]
    \item 	The censoring probability for missing due to other reasons is $p(C_{(i,other)}=1 | L_i,A_i,C_{(i,insti)}=0,D_{(i,death)}=0,D_{(i,ill)}=0)$. We estimated this probability by logistic regression
\[
\text{logit}(P(C_{other}=1 | L,A,C_{insti}=0,D_{death}=0,D_{ill}=0))=\beta_0 + \vec{\beta} \cdot \vec{L},
\]
where $\vec{L}$ is a vector of the baseline covariates: age, race, performance status, and baseline QoL score, $\beta_0$ is the intercept, and $\vec{\beta}$ is a vector of coefficients.

\item  	As a consequence, for a survivor $i$ who is not ill with a non-missing 12-month QoL score, the Inverse Probability of Censoring Weight is

\[ 
\hat{w}_i = \frac{\mathbbm{1}_{(C_{i,insti} = 0, C_{i,other} = 0, D_{i,death} = 0, D_{i,ill} = 0)}}{\hat{p}_{IPCW}},
\]
where
\begin{align*}
\hat{p}_{IPCW} = &\hat{p}(C_{i,other} = 0 | L_i, A_i, C_{i,insti} = 0, D_{i,death} = 0, D_{i,ill} = 0) \times \\
    & \qquad \hat{p}(C_{i,insti} = 0 \mid A_i, D_{i,death} = 0, D_{i,ill} = 0).
\end{align*}

\item  The Inverse Probability of Censoring Weight is 1 for those who died or have missing scores due to illness, since their data are fully observed after data preprocessing and they do not need to account for others’ missing scores.

\end{enumerate}

IPCW assigns the weights above to subjects who were alive and reported their QoL scores, and to subjects who died or were ill with an assigned QoL score. In our application, the sum of the weights was 306.1 for $A=0$ and 315.2 for $A=1$, which are close to the number of subjects who reported their QoL score at baseline: 306 for $A=0$ and 315 for $A=1$, suggesting that such weights for IPCW were successfully assigned.

\subsection{Weighted estimation of the survival-incorporated median}

After preprocessing the data and calculating the weights for IPCW, we estimate the survival-incorporated median using IPCW. We used the following formula, which is a reweighted version of the quantile estimation procedure \citep{koenker1978regression, firpo2007efficient}, to estimate the survival-incorporated median in each treatment arm $a$ separately:
\begin{align*}
    \hat{q}_{(0.5, a)} = \argmin_q \sum^{N_a}_{i=1} \hat{w}_{i} \cdot \rho_{\tau} (Y_i -q), 
\end{align*}

where $\hat{q}_{(0.5, a)}$ is the estimated survival-incorporated median under treatment $a$, $a=0,1$.  $N_a$ is the number of subjects who received treatment $a$. $Y_i$ is the month-12 QoL score if subject $i$ is alive with a non-missing month-12 QoL score, or the assigned score if subject $i$ died (-10) or was ill (-5). $\rho_\tau (Y_i - q)$ is a quantile loss function: $\rho_\tau (Y_i-q)=(Y_i-q)(\tau-\mathbbm{1}_{Y_i-q<0})$ . As mentioned in the last subsection, $\hat{w}_i$, the estimated weight, has the following form:
\[
\hat{w}_i = \begin{cases}
  \frac{\mathbbm{1}_{(C_{i,insti} = 0, C_{i,other} = 0, D_{i,death} = 0, D_{i,ill} = 0)}}{\hat{p}_{IPCW}}  \text{ if $D_{i, death} = 0$ or $D_{i, ill} = 0$}\\
  1 \text{ if $D_{i, death} = 1$ or $D_{i, ill} = 1$},
\end{cases}
\]
\begin{align*}
\hat{p}_{IPCW} = &\hat{p}(C_{i,other} = 0 | L_i, A_i, C_{i,insti} = 0, D_{i,death} = 0, D_{i,ill} = 0) \times \\
    & \qquad \hat{p}(C_{i,insti} = 0 \mid A_i, D_{i,death} = 0, D_{i,ill} = 0).
\end{align*}

We used the R function “weighted\_quantile” from the R package “MetricsWeighted” to estimate the survival-incorporated median QoL score under each treatment separately. We computed the confidence interval for quantile estimators with IPCW using bootstrapping \citep{efron1994introduction}.

\section{The survival-incorporated 75th quantile versus SACE in SWOG S9916} \label{A.sec4}

\citet{ding2011identifiability} proposed an estimation method for the Survival Average Causal Effect (SACE), and they also used the SWOG S9916 data for illustration. They estimated the SACE of MP versus DE on the month-12 change of QoL scores from baseline. The data used in \citet{ding2011identifiability}, which include 487 subjects of the total 674 subjects, are a subset of original data. Because all survivors with missing month-12 QoL scores were excluded, compared with the original data, the probability of death in this data subset is substantially altered. Supplementary Table \ref{tab:A.1} summarizes the number of subjects and the probability of death under each treatment in this data subset. For a fair comparison of SACE with survival-incorporated quantiles, here we use the same data subset that was used in \citet{ding2011identifiability}.

\begin{table}[h] 
\centering
\renewcommand{\arraystretch}{1.2}
\begin{tabular}{|c c c|}
\hline
& MP ($A=0$) & DE ($A=1$) \\
\hline
\#subjects & 229 & 258 \\
\%death at 12 months & 140 (61.1\%) & 130 (50.39\%) \\
\hline

\end{tabular}
\caption{Number of subjects and the probability of death under each treatment in the data subset with 487 subjects. MP: treatment of mitoxantrone and prednisone. DE: treatment of docetaxel and estramustine.}
\label{tab:A.1}
\end{table}

\begin{table}[h] 
\centering
\renewcommand{\arraystretch}{1.4}
\begin{tabular}{|l|c|c|c|}
\hline
\makecell{Estimates \\ (95\% Confidence Interval)} & MP ($A=0$) & DE ($A=1$) & Effect estimate \\
\hline
\makecell{Survival-incorporated \\75th quantile QoL change} & \makecell{-16.7 \\(-31.8, -1.6)} & \makecell{0 \\(-14.2, 14.2)} & \makecell{16.7 \\(-4.1, 37.4)} \\
\hline

\makecell{SACE of QoL Change * \\ (monotonicity)} & \makecell{-9.1 \\(-14.1, -4.6)} & \makecell{-2.1 \\(-31.9, 18.2)} & \makecell{7.0 \\(-23.9, 26.1)} \\
\hline
\makecell{SACE of QoL Change \\ (no monotonicity)} & \makecell{-19.1 \\(-24.0, -12.3)} & \makecell{-19.7 \\(-23.2, -12.6)} & \makecell{0.6 \\ (-11.3, 10.2)} \\
\hline
\end{tabular}
\caption{The estimated survival-incorporated median 75th quantile QoL change at month 12 and the SACE of QoL change at month 12 with and without monotonicity. QoL score: Quality of Life score. MP: the treatment of mitoxantrone and prednisone. DE: the treatment of docetaxel and estramustine. The analysis was performed on the SWOG S9916 data subset (487 subjects).}
\label{tab:A.2}
\end{table}

We estimated the survival-incorporated 75th quantile of the month-12 change of QoL scores in each treatment arm, because the probability of death in the resulting dataset is greater than 50\%. As mentioned in the Discussion, since the clinical outcome of interest is the change in QoL score, to estimate the survival-incorporated 75th quantile, we assigned patients who died a value (-200) less than the lowest value of the change of QoL scores. In this data subset, there are no missing month-12 QoL scores. Thus, instead of using IPCW, we simply computed the sample 75th quantile in each treatment arm, incorporating patients who died. We estimated the SACE using the methods described in \citet{ding2011identifiability}, and we used the bootstrap method to construct confidence intervals for the estimated SACEs. 

Supplementary Table \ref{tab:A.2} shows the estimated survival-incorporated 75th quantile, as well as the estimated SACE, with and without assuming monotonicity. These summary measures are better under DE than under MP, although the confidence intervals of the estimated effects all include zero. The effect based on the survival-incorporated 75th quantile (16.7) is greater than the effect based on the SACE with monotonicity (7.0) and without monotonicity (0.6). Compared to the estimation procedure for the survival-incorporated median, the estimation procedure for the SACE is more complex even with monotonicity. However, the monotonicity assumption is likely not valid in the SWOG S9916 study, which compares two active treatments: MP and DE. In contrast, the survival-incorporated 75th quantile, which relies on no assumptions if no missing data, is easy to compute and can be a useful summary measure when a substantial number of patients die.
\end{appendices}

\end{document}